\documentclass[10pt]{article}
\usepackage{verbatim}
\usepackage{graphicx}
\usepackage{times}
\usepackage{amsmath,amsfonts,amssymb,latexsym}
\usepackage{mathpazo}
\PassOptionsToPackage{hyphens}{url}

\usepackage{listings, color}    
\definecolor{dkgreen}{rgb}{0,0.6,0}
\definecolor{gray}{rgb}{0.5,0.5,0.5}
\definecolor{mauve}{rgb}{0.58,0,0.82}
\definecolor{blue}{rgb}{0.13, 0.58, 0.99}
\definecolor{black}{rgb}{0,0,0}
\definecolor{darkgray}{rgb}{0.3, 0.3, 0.3}
\definecolor{red}{rgb}{0.8, 0.2, 0.2}

\lstdefinelanguage{JavaScript}{% 
  keywords={typeof, new, true, false, catch, function, modifier, contract, return, null, catch, switch, var, if, in, while, do, else, case, break, throw},
  keywordstyle=\color{blue}\bfseries,
  ndkeywords={class, export, boolean, implements, import, this, uint32, uint256, mapping, address, public},
  ndkeywordstyle=\color{darkgray}\bfseries,
  identifierstyle=\color{black},
  sensitive=false,
  comment=[l]{//},
  morecomment=[s]{/*}{*/},
  commentstyle=\color{dkgreen}\ttfamily,
  stringstyle=\color{red}\ttfamily,
  morestring=[b]',
  morestring=[b]",
  basicstyle=\footnotesize\ttfamily
}

\usepackage[justification=centering]{caption} % to center the captions

\usepackage{color}
\usepackage[usenames,dvipsnames,svgnames,table]{xcolor}
\usepackage[colorlinks=true,
            linkcolor=gray,
            urlcolor=blue,
            citecolor=red]{hyperref}

\usepackage[margin=0.5in,bmargin=2cm]{geometry}

\usepackage{multicol}

\begin{document}
\title{%{\small [draft]}\\
{\bf The Blockchain Anomaly}}

\author{Christopher Natoli\\{\small NICTA/Data61-CSIRO}\\{\small University of Sydney}\\\small{cnat5672@uni.sydney.edu.au}
\and Vincent Gramoli\\{\small NICTA/Data61-CSIRO}\\{\small University of Sydney}\\\small{vincent.gramoli@sydney.edu.au}}
%\author{\IEEEauthorblockN{Christopher Natoli}
%\IEEEauthorblockA{\small NICTA/Data61-CSIRO\\
%University of Sydney\\
%\small{cnat5672@uni.sydney.edu.au} }%\vspace{-2em}}
%\and
%\IEEEauthorblockN{Vincent Gramoli}
%\IEEEauthorblockA{\small NICTA/Data61-CSIRO\\
%University of Sydney\\
%\small{vincent.gramoli@sydney.edu.au}}}%\vspace{-2em}}

\date{}

\maketitle

\thispagestyle{plain}
\pagestyle{plain}

\begin{abstract}
Most popular blockchain solutions, like Bitcoin, rely on proof-of-work, guaranteeing that the output of the consensus is agreed upon with high probability. However, this probability depends on the delivery of messages and that the computational power of the system is sufficiently scattered among pools of nodes in the network so that no pool can mine more blocks faster than the crowd. New approaches, like Ethereum, generalise the proof-of-work approach by letting individuals deploy their own private blockchain with high transaction throughput. As companies are starting to deploy private chains, it has become crucial to better understand the guarantees blockchains offer in such a small and controlled environment.

In this paper, we present the \emph{Blockchain Anomaly}, an execution that we experienced when building our private chain at NICTA/Data61. 
Even though this anomaly has never been acknowledged before, it may translate into dramatic consequences for the user of blockchains.
Named after the infamous Paxos anomaly, this anomaly makes dependent transactions, like ``Bob sends money to Carole after he received money from Alice'' impossible. This anomaly relies on the fact that existing blockchains do not ensure consensus safety deterministically: 
there is no way for Bob to make sure that Alice actually sent him coins without Bob using an external mechanism, like converting these coins into a fiat currency that allows him to withdraw. 
We also explore smart contracts as a potential alternative to transactions in order to freeze coins, and show implementations of smart contract that can suffer from the Blockchain anomaly and others that may cope with it. \\

\end{abstract}

%\begin{multicols}{2}
\twocolumn

\section{Introduction}

\sloppy{Mainstream public blockchain systems, like Bitcoin~\cite{Nak08} and Ethereum~\cite{Woo14}, require to reach consensus on Internet despite the presence of malicious participants.
Yet, it is impossible for a distributed system including a faulty participant to reach consensus if messages may not be delivered within a bounded time~\cite{FLP85}.
This contradiction raises interesting research questions regarding the formal properties that are sacrificed in these blockchain systems.
%It is impossible for a distributed system including a faulty process to reach consensus if messages cannot be delivered within a bounded time.
%It is impossible for a distributed system including a faulty process to reach consensus if messages cannot be delivered within a bounded time.
%Yet, predominant blockchain systems, like Bitcoin~\cite{Nak08} and Ethereum~\cite{Woo14}, require to reach consensus on Internet despite the presence of malicious participants.
Foundational consensus algorithms~\cite{DLS88} were proposed to never reach a decision in case of arbitrary message delays, but to respond only correctly if ever.
Surprisingly, %we observed that 
these blockchain systems adopt a different approach, sometimes responding incorrectly.
These few last years, the concept of \emph{private chain} gained traction for its ability to offer blockchain among multiple companies 
in a private, controlled environment. 
Three months ago, eleven banks collaborated successfully in deploying an Ethereum private chain to perform transactions across North America, Europe and Asia.\footnote{BMO Financial Group, Credit Suisse, CBA, HSBC, Natixis, Royal Bank of Scotland, TD Bank, UBS, UniCredit and Wells Fargo as explained at \url{http://www.ibtimes.co.uk/r3-connects-11-banks-distributed-ledger-using-ethereum-microsoft-azure-1539044}.}
To understand the limitations of consensus and its potential consequences in the context of private chains,  
we deployed our own private chain and stress-tested the systems in corner-case situations.}
%, and simulated a network attacker delaying messages. 

%In this paper, we build upon our experience of deploying a private blockchain and we list our observations.
%In particular, 
In this paper, we present the  \emph{Blockchain anomaly},  a new problem named after the Paxos anomaly~\cite{HKJ+10,BMvR10,BMvR12,GBFS15}, that 
prevents Bob from executing a transaction based on the current state of the blockchain.
In particular, we identified a complex scenario where the agreement on the state of the blockchain
is not sufficient to guarantee immutability of the chain.
This anomaly can lead to dramatic consequences, like the loss of virtual assets or a double-spending attack.
%To go further, we identify the \emph{Blockchain anomaly}, a new problem called after the Paxos anomaly, that 
%prevents Bob from executing a transaction based on the current state of the blockchain.
We also show that some \emph{smart contracts}, expressive code snippets that help defining how virtual assets can be owned and exchanged in the system, 
may suffer from the Blockchain anomaly.
Our results outline the risk of using a blockchain in a private context without understanding its complex design features.
We terminate our experience report by providing the source code of a more complex smart contract that can circumvent a particular 
example of the Blockchain anomaly.
%We show reproducible
%executions where agreed transactions can disappear from the chain.
%Finally, we present a novel problem, called the \emph{Blockchain anomaly}, making it impossible to execute a transaction based on 
%current state of the system.

Most blockchain systems track a transaction by including it in a block that gets mined 
%by solving a computationally hard problem 
before being appended 
to the chain of existing blocks, hence called \emph{blockchain}.
The consensus algorithm guarantees a total order on these blocks, so that the chain does not end up being a tree.
This process is actually executed speculatively in that multiple new blocks can be appended transiently to the last block of the chain---a transient branching process known as a \emph{fork}.
Once the fork is discovered, meaning that the participants learn about the two branches, the 
longest branch is adopted as the valid one.
%new blocks get appended to the longest branch.
%and the other branch (including the transactions of its blocks) get either discarded.
Blockchain systems usually assume that forks can grow up to some limited depth, as extending a branch requires to solve a complex challenge that boils down to spending 
a long time during which one gets likely notified of the longest chain.
Bitcoin recommends six blocks to be mined after a transaction is issued to consider the transaction accepted by the system.
%\vincent{Is it 36 blocks in homestead?} 
Similarly, Ethereum states that five to eleven more blocks should be appended after a block for it to be accepted~\cite{Woo14}.
%Ethereum indicates that all transactions of a block with 5 successor blocks in the chain is ``validated'' indicating that there cannot exist a longer chain~\cite{Woo14}.
%\footnote{Actually, the long-range attack could leverage the turing complete nature of the smart contract code to discard a chain by producing a longer one as described at \url{https://blog.ethereum.org/2014/05/15/long-range-attacks-the-serious-problem-with-adaptive-proof-of-work/}.}

However, consensus cannot be solved in the general case. In particular, 
foundational results of distributed computing indicate that consensus cannot be reached if there is no upper-bound on the time for a message to be delivered and if some participant may fail~\cite{FLP85}.
Consensus is usually expressed with three properties: \emph{agreement} indicating that if two non-faulty participants decide they decide on the same block, \emph{validity} indicating that the decided block should be one of the blocks that 
were proposed and \emph{termination} indicating that eventually a correct participant decides.
The common decision that is taken by famous consensus protocols, like Paxos~\cite{Lam98} and Raft~\cite{OO14},
%, Zyzzyva?, Aleph?, PBFT? 
is to make sure that if the messages get delayed, at least validity and agreement remain ensured. This is achieved by having the algorithm doing nothing in the worst case, hence sacrificing termination to ensure that only correct responses---satisfying both validity and agreement---can be returned.
These consensus algorithms are appealing, because if after some time the network stabilises and messages get delivered in a bounded time, then consensus is reached~\cite{DLS88}.

We illustrate the Blockchain anomaly and describe a distributed execution where even committed transactions of a private chain get reordered so that the latest transaction ends up being committed first. We chose Ethereum for our experiments as it is a mainstream blockchain system that allows the deployment of private chains.
We show how to reproduce the Blockchain anomaly by following the same execution, where messages get 
delayed between machines while some miner mines new blocks. Despite transactions being already committed the eventual delivery of messages produces a reorganisation reordering some of the committed transactions. 
In our execution, miners are setup to dedicate different number of cores to the mining process, hence mining at different speeds. We argue that the misconfiguration of a machine and the heterogeneous mining capabilities of machines belonging to different companies are sufficiently realistic to allow an attacker to execute a double-spending attack. %make the resulting problem interesting.
Finally, we discuss the relations of the Blockchain anomaly to other problems and 
observe that it is not confined to 
the Ethereum blockchain but could potentially apply to proof-of-stake private blockchains as well, 
requiring further investigations.

%In this paper, we show that blockchain systems, by adopting a different approaches, make it hard for their users to transfer digital assets safely.
%% in that they do not sacrifice termination, but rather sacrifice agreement.
%We show that this problem can have dramatic effects like the definitive unwanted loss of digital assets or a double-spending attack.
%To this end, we deployed a private Ethereum blockchain on a controlled network of machines and simulated a network attack
%% can DoS result in this issue?
%that delayed the communication among participants.
%%
%We present a new problem, the \emph{Blockchain anomaly}, called after the Paxos anomaly~\cite{HKJ+10,BMvR10,BMvR12} and describe reproducible executions that lead to this anomaly.
%The result is the violation of the safety of consensus.
%We conclude by explaining how carefully written smart contracts can help 
%replacing transactions that suffer from the Blockchain anomaly.
%should be done by non-expert blockchain to limit these scenarios using smart contracts.

Section~\ref{sec:prel} overviews the blockchain technology, the Paxos anomaly and defines the important terms of the paper.
In Section~\ref{sec:ba}, we present the blockchain anomaly.
In Section~\ref{sec:eval}, we present our experiments based on an Ethereum private chain.
In Section~\ref{sec:sc}, we explain how replacing transactions by smart contracts could help bypassing the anomaly.
Section~\ref{sec:disc} discusses the Blockchain anomaly in other settings.
Section~\ref{sec:rw} presents the related work.
And Section~\ref{sec:conclusion} concludes.

\section{Preliminaries}\label{sec:prel}

In this section, we present the key concepts of Bitcoin and Ethereum consensus protocols, the condition of their 
termination and the Paxos anomaly before presenting the general model.

\subsection{Blockchain Systems}
\begin{figure}
\begin{center}
\includegraphics[scale=0.42, clip=true, viewport=50  300 700 550]{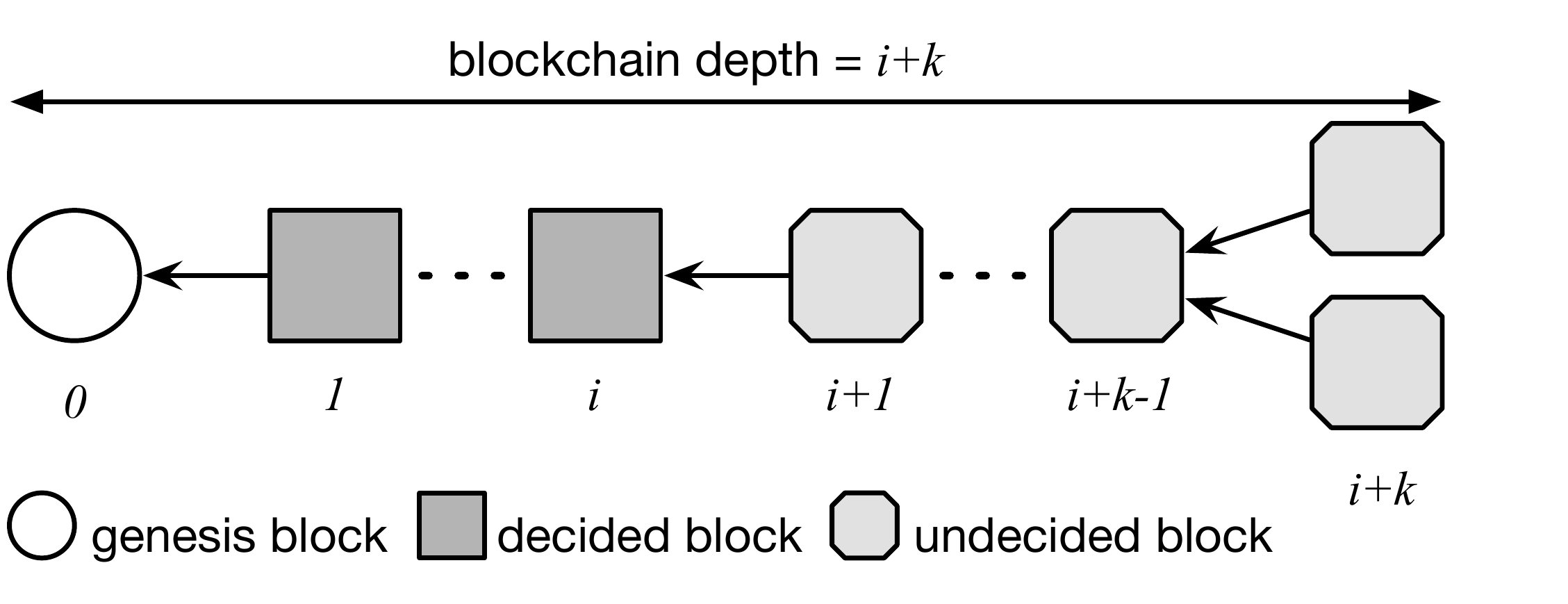}
\caption{The blockchain structure starts with a genesis block at index $0$ and links successive blocks in reverse order of their index; a new block is decided at index $i>0$ when the blockchain depth reaches $i+k$ (note that a blockchain of depth 0 is a genesis block)\label{fig:termination}}
\end{center}
\end{figure}

A blockchain can be considered as a replicated state machine~\cite{XPZ16} where a reversed link between blocks 
%represents a transition between states~\cite{XPZ16} as depicted in Figure~\ref{fig:termination}.
is a pointer from a state to its preceding state as depicted in Figure~\ref{fig:termination}.
Consensus is necessary to totally order the blocks, hence maintaining the chain structure.
To reach consensus despite arbitrary failures, including malicious behaviors, traditional blockchain systems adopted a technique based on proof-of-work, requiring a proof of computation~\cite{DN92}.
Specialised peers, called \emph{miners}, provably solve a hashcash crypto puzzle~\cite{Bla02} before a new block can be appended to 
the blockchain.
%the network of participants can accept a new block of transactions.
Given a block and a threshold, a miner repeatedly selects a nonce and applies a pseudo-random function to this block and the selected nonce until it obtains a result lower than the threshold.
The difficulty of this work limits the rate at which new blocks can be generated by the network.

%Proof-of-work~\cite{DN92} was originally applied to the context of consensus to avoid the possibility of having processes joining the system under multiple identities.

\subsection{From Nakamoto's Consensus to Smart Contracts}
% nakamoto's consensus
Nakamoto's consensus~\cite{Nak08} is at the core of Bitcoin, the mainstream decentralised digital currency. % and can be expressed in terms of agreement, validity and termination.
%as follows where $N$ is the set of all nodes and $B(t)\subset N$ is the set of Byzantine nodes at time $t$ and the mining power of node $i \in N$ is denoted $m(i)$.
%\begin{itemize}
%\item {\bf N-agreement:} There exists a time difference function $\Delta(.)$ such that, given $0<\epsilon<1$, the probability that at time $t$ two nodes return the same state for $t-\Delta(\epsilon)$ is higher than $1-\epsilon$.
%\item {\bf N-validity:} If the fraction of mining power of Byzantine nodes is strictly upper-bounded by $f$, then the average fraction of state machine transitions that are not inputs of honest nodes is smaller than $f$.
%\item {\bf N-termination:} There exists a time difference function $\Delta(.)$ such that, given a time $t$ and a value $0<\epsilon<1$, the probability that for two times $t',t'' > t+\Delta(\epsilon)$, a node returns two different states for the machine at time $t$ is higher than  $1-\epsilon$.
%\end{itemize}
Interestingly, Nakamoto's consensus does not guarantee 
%termination and 
agreement deterministically. Instead it guarantees that 
%termination and agreement are met 
agreement is met with some probability close to 1.
%Bitcoin relies on the concept of proof-of-work, requiring specialized nodes, called \emph{miners}, to solve a hashcash crypto puzzle before the network of participants can accept a new block of transactions.
%Given a block and a threshold, a miner repeatedly selects a nonce and applies a pseudo-random function to this block and the selected nonce until it obtains a result lower than the threshold.
%The difficulty of this work limits the rate at which new blocks can be generated by the network, the rate in 
The difficulty of the crypto puzzles used in Bitcoin leads to mining a block every 10 minutes.
The advantage of this long period, is that it is relatively rare for the blockchain to \emph{fork} due to blocks being simultaneously mined and Bitcoin resolves these forks by choosing the longest branch and discarding the other(s). 

Ethereum~\cite{Woo15} is a recent open source cryptocurrency platform that also builds upon proof-of-work. As opposed to Bitcoin's consensus protocol, Ethereum generates one block every 12--15 seconds. While it improves the throughput (transactions per second) it also favours transient forks as miners are more likely to propose new blocks simultaneously. 
To avoid frequently wasting mining efforts to resolve forks, Ethereum uses the {\sc Ghost} (Greedy Heaviest Observed Subtree) protocol that does not necessarily discards all the, so called uncle, blocks of non selected branches.  
Ethereum offers a Turing-complete programming language that can be used to write \emph{smart contracts}~\cite{Sza97} that define new ownership rules.
%\vincent{As Ethereum contracts can include any kind of computation, Ethereum prevents potentially ASICs (application-specific integrated circuits) from mining faster than regular general 
%purpose processing unit.}
% 
%Ethereum is a generalisation of the underlying concept of Bitcoin 
%in that it offers the possibility for the user to write a \emph{smart contract}, a program snippet that can be triggered when a transaction is issued and that 
%can, in response, trigger other smart contracts or transactions.
%Ethereum uses miners like Bitcoin and \emph{peers} that are clients that issue transactions to the system but do not mine.
%
%\vincent{Transactions are included into blocks that are mined before being added to the chain. Ethereum considers that
%consensus terminates ($b$ is decided) whenever 12 new blocks are appended after $b$}.
% that a block $b$ is safely inserted once 5 new blocks were appended after $b$.
%
%Strato has recently announced a partnership with Microsoft to make Ethereum available on the Azure cloud environment.\footnote{\url{http://blockapps.net/pdfs/blockapps-strato-microsoft-partnership.pdf}.}

\subsection{Termination of Consensus}\label{ssec:termination}
By relaxing the agreement property of consensus, blockchain systems can guarantee termination deterministically. In the context of 
blockchain, termination of consensus indicates that a block has been \emph{decided} for the next available block index.
We say that all the transactions of a decided block are \emph{committed}.\footnote{Here, we use the term ``committed'' rather than ``confirmed'' as, in the blockchain terminology, a transaction is  meant to be ``confirmed'' sometimes when only its block is mined, and sometimes when $k+1$ blocks get mined (its own block and the $k$ successor blocks).}
This decision upon a block inclusion in the chain is necessary for cryptocurrency exchange platforms, for example, to determine that coins of a particular type that are newly minted\footnote{As opposed to \emph{mining} that includes the computation of the miners, \emph{minting} consists simply of the creation of coins.} within this block can be converted into \emph{altcoins} (coins of a different type) or fiat currencies (e.g., EUR, USD). In particular, observing that a block was mined and appended to the chain is not sufficient to guarantee that it is decided: this block could be part of one branch of a transient fork without consensus being reached yet on any of these branches.

Figure~\ref{fig:termination} depicts the termination of consensus on the index $i$ of a blockchain. An arrow pointing from left to right indicates that a block contains a hash of its predecessor block, the one located immediately on its left. Newly mined blocks are added to the right end of the blockchain that may fork transiently if multiple blocks referring to the same predecessor get mined concurrently. Forks are only transient and their resolution depends on the blockchain system in use. The consensus for an index $i$ terminates when participants decide on the new block to be assigned at index $i$. The decision upon the block at index $i$ occurs for all $i>0$ when the blockchain depth reaches $i+k$, where $k\geq 0$ is a constant dependent on the Blockchain system.

Different blockchain systems adopt different values of $k$ to define termination.
In Bitcoin (btc), $k_{btc}=5$, meaning that the block at index $i$ is decided---consensus for index $i$ terminates---when the $k_{btc}+1=6$ blocks at indices $i, ..., i+5$ have been successfully mined. As we previously mentioned, a new block is decided every 10 minutes in Bitcoin, hence it takes $(k_{btc}+1)*10\,min = 1\,hour$ for a transaction to be committed in Bitcoin.
%
%The drawback, however, is that a transaction is not decided (consensus is not reached) before 
%the $(i+5)^{th}$  block gets mined with $i$ the index of the block containing this transaction,
%meaning that the block and 5 subsequent blocks need to be mined by the system. Hence termination as considered by typical wallets takes one hour ($6 \times 10$ minutes).\footnote{\url{https://en.bitcoin.it/wiki/Confirmation}.} \vincent{Replace the 6 confirmations by another term as it is sometimes considered to be 1 confirmation.}
%
In Ethereum (eth) since verion 1.3.5 Homestead, $k_{eth}=11$, meaning that the block at index $i$ is decided---consensus for index $i$ terminates---when the blockchain depth reaches $i+11$. Hence it takes $(k_{eth}+1)*15\,sec = 3\,min$ for transactions to be committed in Ethereum.
Note that some cryptocurrency exchange platforms adopt different values of $k$ to adjust the probability of agreement, hence QuadrigaCX Ether Trading waits for $k_{btc}'+1=4$ blocks to be mined in the Bitcoin blockchain while it waits for $k_{eth}+1=12$ blocks   
to be mined in the Ethereum blockchain.\footnote{\url{https://www.quadrigacx.com/faq}.}
%Ethereum, a transaction present in the $i^{th}$ block is considered committed after the $i+11$ block gets appended to the chain.
%Although this number has changed with the development progresses of Ethereum, the Mist Ethereum user interface currently adopted it.
%Trading systems, like QuadrigaCX Ether Trading waits for $k^{qcx}_{btc}=4$ confirmations in Bitcoin and $k^{qcx}_{ether}=12$.\footnote{\url{https://www.quadrigacx.com/faq}.}
%\vincent{LocalBitcoins $3 \leq k^{lcb}_{btc} \leq 6$.}

\subsection{The Paxos Anomaly}
\begin{figure}[t]
\begin{center}
\includegraphics[clip=true, viewport=30 280 450 505, scale=0.5]{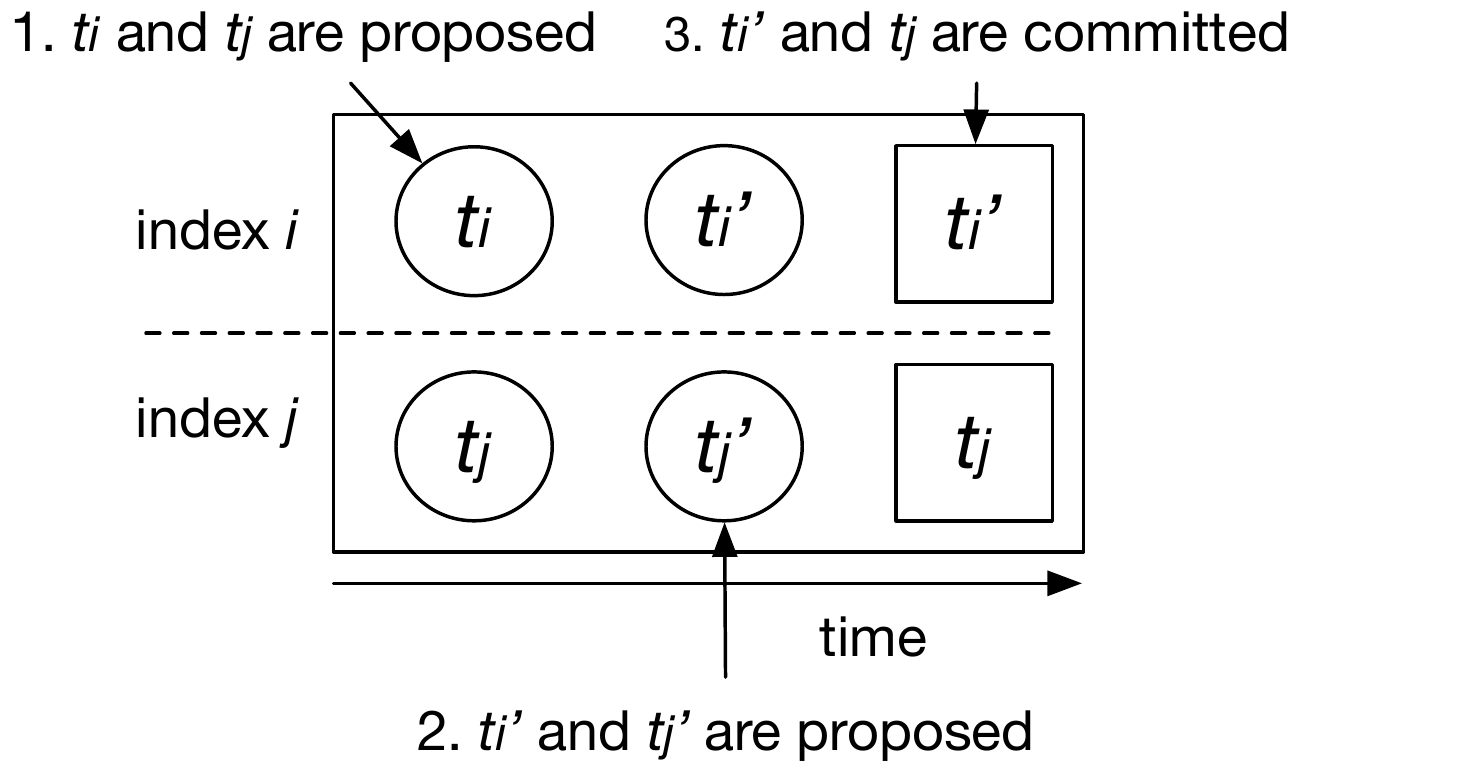}%\vspace{-15em}
\caption{The Paxos anomaly: a first leader proposes $t_i$ and $t_j$ for slots $i$ and $j > i$ (with $t_j$ being implicitly conditional to the commit of $t_i$), a second leader proposes $t_i'$ and $t_j'$ while a third leader commits $t_i'$ and $t_j$ for slots $i$ and $j$, respectively, hence violating the dependency between $t_i$ and $t_j$\label{fig:pa}}
\end{center}
\end{figure}

%An appealing feature for designing cluster membership protocols in a centralized way is avoiding the \emph{Paxos anomaly}~\cite{BMvR10,HKJ+10}:
Paxos is a famous consensus protocol originally guaranteeing agreement and validity despite crash failures~\cite{Lam98}.
The \emph{Paxos anomaly}~\cite{BMvR10,HKJ+10} stems from the difficulty of implementing conditional requests (or transactions) in Paxos:
Paxos decides on individual proposed transactions, potentially violating dependencies between transactions even when proposed by the same requester as depicted in Figure~\ref{fig:pa} where a 
slot can be viewed as the index of the decision.
These dependencies can be useful to make the execution of a transaction $t_j$ dependent on the successful execution of a previous transaction $t_i$: for example if Bob wants to transfer an amount of money to Carole ($t_j$) only if he successfully received some 
money from Alice ($t_i$).
In centralised systems, this anomaly can be easily avoided by enforcing an ordering on these transactions by simply forwarding all requests to a primary node or coordinator~\cite{HKJ+10}.
%update operations can partially update a tree structure of znodes hence clients can request to create a znode and request, immediately after, to create a child node of this parent znode. If these two requests are concurrent, Paxos can decide upon one of the two operations, say the creation of the child node, and discards the other, say the creation of the parent znode. This would lead to an incorrect state where the child node cannot be created because its parent znode does not exist.
In Paxos as in fully decentralised systems, however, the first transaction may not be decided in favour of another proposed transaction in a first consensus instance while in a subsequent consensus instance the second transaction may be successfully decided.
This results in a violation of the condition that the second transaction should be decided only if the first transaction was decided.

Below we present the Blockchain anomaly due to the decentralised aspects of blockchain systems, 
like Bitcoin and Ethereum. The Blockchain anomaly shares similarities with the Paxos anomaly, except that it can occur when
transactions, issued by different nodes of the system, are not even concurrent.

%Paxos decides on individual proposed values, potentially violating  dependencies that requesters have between their proposed values.

\subsection{General Model}

We consider a distributed blockchain system of $n$ peers where peers can exchange coins from one to another 
through \emph{transactions}.
The goal is for the system to implement a \emph{ledger} abstraction as a public permanent
and auditable records of all transactions.
The ledger is implemented with a \emph{blockchain}, a series of transaction blocks, starting with a special block called the \emph{genesis block}.
Blocks are singly linked one after another up to the genesis block---each non-genesis block containing a hash of the previously accepted block---and 
define the current state of the ledger as the set of transactions that ever occurred.
The block index or \emph{slot} increases monotonically from the index 0 of the genesis block. 

Peers can fail arbitrarily, they can stop working and can be malicious.
Any peer can issue transactions that get recorded into the \emph{transaction pool}. 
Only miners can bundle a subset of the pool of transactions into a block after ensuring that there are
sufficient funds available on the accounts of the ledger and that these transactions do not conflict.
%This \emph{mining} process requires the miner to solve a crypto puzzle.
The system uses consensus to guarantee that no malicious peers are trying to double-spend some coins by issuing two 
conflicting transactions concurrently to different miners. 
To this end, we consider a consensus protocol based on proof-of-work so that miners bundle transactions from the transaction pool into a block by 
solving a crypto puzzle in exchange of coins for the system to decide on the new block. 
%Each miner takes transactions from the transaction pool and mines a block containing non-conflicting transactions by looking for the solution of the puzzle.
%Whenever the solution is found by a miner that has mined a block, the block is appended to the main chain. 
The \emph{difficulty} of the crypto puzzle determines the rate at which new blocks can be mined: the higher the difficulty, the slower the rate.

At times though the blockchain may fork transiently\footnote{This transient forks are sometimes referred to as \emph{ephemeral forks} as they are expected in the normal course of the execution.}, indicating 
that multiple blocks were appended to a unique block, in which case conflicting transactions could potentially be part of blocks on 
different branches of the chain.
If this fork happens, 
%multiple miners have appended different blocks to the main chain concurrently, 
then a \emph{reorganisation} 
process eventually occurs to resolve the fork by uniquely identifying one of the two branches as the right one.

%Proof-of-work blockchains adopt a parameterised termination, that we call $k$-termination:
%whenever $k$ blocks are confirmed by the system, then the first of these block is considered \emph{decided}, meaning that consensus for this particular block is reached. 
%Note that even though the $k-1$ last blocks are confirmed they are not decided yet. We say that the transactions of the decided block are \emph{committed}.
%Whenever a subchain of five blocks gets appended to the block $b$, then consensus terminates and the block is decided.

%\section{Problem}
\section{The Blockchain Anomaly}\label{sec:ba}
We present the Blockchain anomaly, an anomaly that affects mainstream blockchain systems whose consensus protocol does not ensures agreement deterministically.
%when used in a permissioned system, like as a private chain. %like \textit{Ethereum}~\cite{Woo14}.

\subsection{Causes of the Blockchain Anomaly}
The problem stems from the asynchrony of the network, in which message delays cannot be bounded, and the termination of consensus.
Although two miners mine on the same chain starting from the same genesis block, a long enough delay in messages between them could lead to having the miners seemingly agree separately on different branches containing more than $k$ blocks each, for any $k$.
This anomaly is dramatic as it can lead to simple attacks within any private network where users have an incentive to maximise their profits---in terms of coins, stock options or arbitrary 
ownership.
Moreover, this scenario is realistic in the context of private chain where the employees of a company, like NICTA/Data61, have direct access to some of the network resources.
When messages get finally delivered, the results of the disagreement creates inconsistencies.
%If the messages between two peers are delayed long enough for a significant chain to be mined with differing information.
%The anomaly occurs due to a long enough message delay between participants for a significant forked chain to be mined with differing information to the original chosen chain. Due to the chain length differences, the policy states that the chain with the most amount of work on it\footnote{The longest chain has most work done.} will be the new chosen chain in the event of a fork. 

\subsection{Uncommitting Transactions}

\begin{figure}[t]
\centering
\includegraphics[clip=true, viewport=70 300 550 530, scale=0.5]{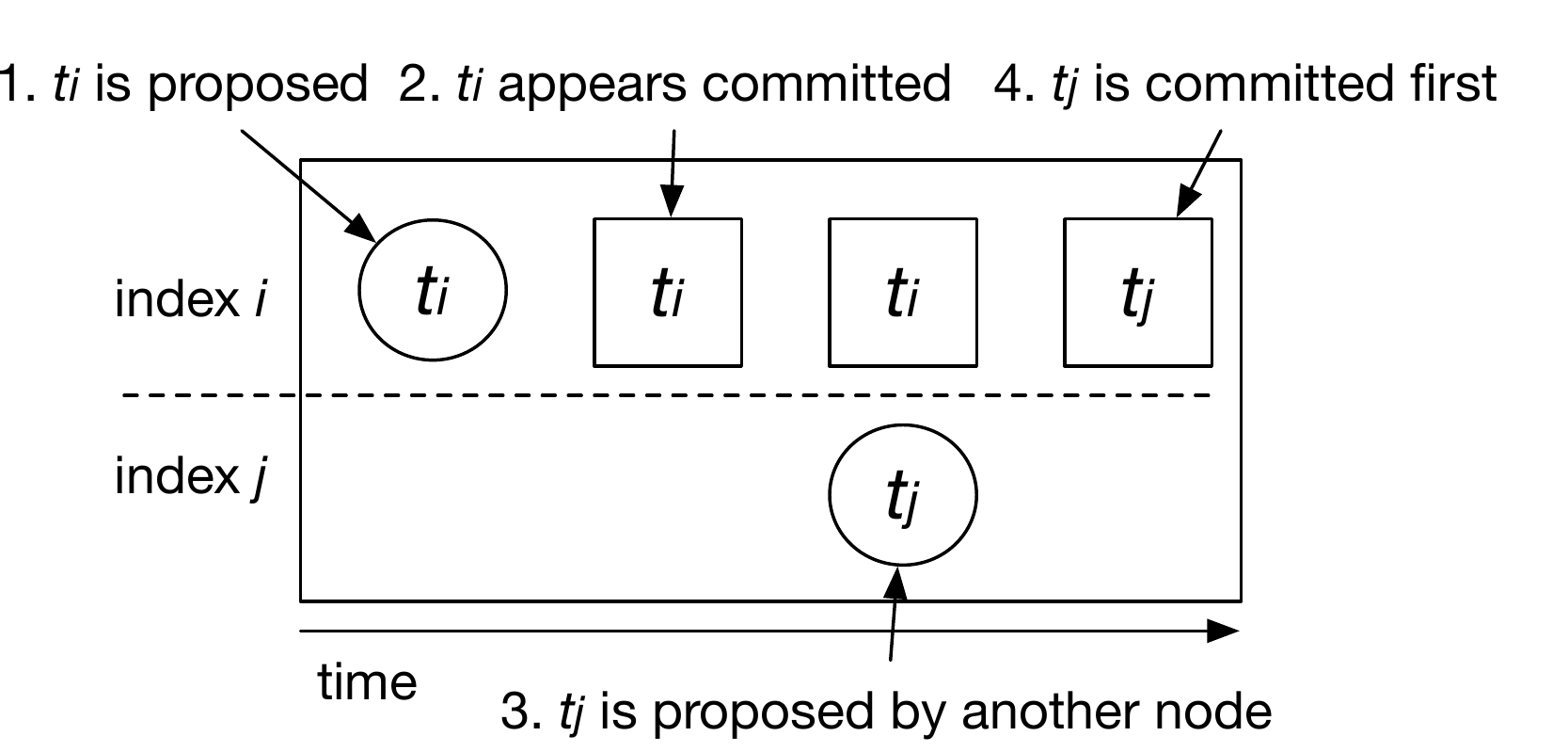}
%\includegraphics[width=0.5\textwidth]{txswap_nochain}  % These are other possibilities of the diagram, please check the  folder
%\includegraphics[width=0.5\textwidth]{txswap}
%\vspace{-15em}
\caption{The Blockchain anomaly: a first client issues $t_i$ that gets successfully mined and committed then a second client issues $t_j$, with $t_j$ being conditional to the commit of $t_i$ (note that $j \geq i+k$ for $t_i$ to be committed before $t_j$ gets issued), but the transaction $t_j$ gets finally reorganised and successfully committed before $t_i$, hence violating the dependency between $t_i$ and $t_j$\label{fig:ba}}
\end{figure}

Figure~\ref{fig:ba} depicts the Blockchain anomaly, where a transaction $t_i$ gets committed as part of slot $i$ from the standpoint of some nodes. Based on this observation, one proposes a new transaction $t_j$ knowing that $t_i$ was successfully committed. 
Again, one can imagine a simple scenario where  ``Bob transfers an amount of money to Carole'' ($t_j$) only if ``Bob had successfully received some 
money from Alice'' ($t_i$) before.
However, once these nodes get notified of another branch of committed transactions, they decide to reorganise the branch to resolve the fork. %As agreement regarding $t_i$ was initially violated, then the transaction $t_i$ does no longer appear as committed in the blockchain. 
The reorganisation removes the committed transaction $t_i$ from slot $i$. Later, the transaction $t_j$ 
%that was proposed based on the observation that $t_i$ 
is successfully committed in slot $i$.

The anomaly stems from the violation of the dependency between $t_j$ and $t_i$: $t_j$ occurred meaning that Bob has transferred an amount of money to Carole, however,  
$t_i$ did not occur meaning that Bob did not receive money from Alice.
Note that in Bitcoin, transaction $t_i$ gets discarded whereas in Ethereum transaction $t_i$ may in some cases be committed in slot $j$.

\subsection{Facilitating a Double-Spending Attack}
One dramatic consequence of the Blockchain anomaly is the possibility for an attacker to execute a \emph{double-spending} attack: converting, for example, all his coins into goods twice.
The scenario consists of the attacker issuing a first transaction $t_1$ that converts all its coins into goods in block $i$ 
and starting mining blocks after block $i-1$ in isolation of the network. 
As part of this mining, the attacker mines another transaction $t_2$ that also converts all its coins into goods.
The attacker then waits for the blockchain depth to reach $i+k$ after what it can collect its goods as a result of transaction $t_1$, 
then it publicises its longer chain without $t_1$ so that the chain gets adopted by the rest of network. 
$t_2$ gets committed in block $j$ and after the chain 
depth reaches $j+k$, the peer can collect its goods for the second time.
% wait for the chain to reach depth $i+k$. Then, the peers start mining  blocks for the transaction to be committed, mine other blocks with a transaction $t_2$ the converts all coins into goods
Note that even if one tries to re-commit $t_1$ later, the transaction will be invalidated because the balance is insufficient, however, the double-spending already occurred.
 
%mined some blocks where he double spend the same coins and finally reorder the two transactions 

\begin{figure*}[!ht]
\centering
\includegraphics[width=1.7\textwidth, clip=true, viewport=-50 200 1500 650]{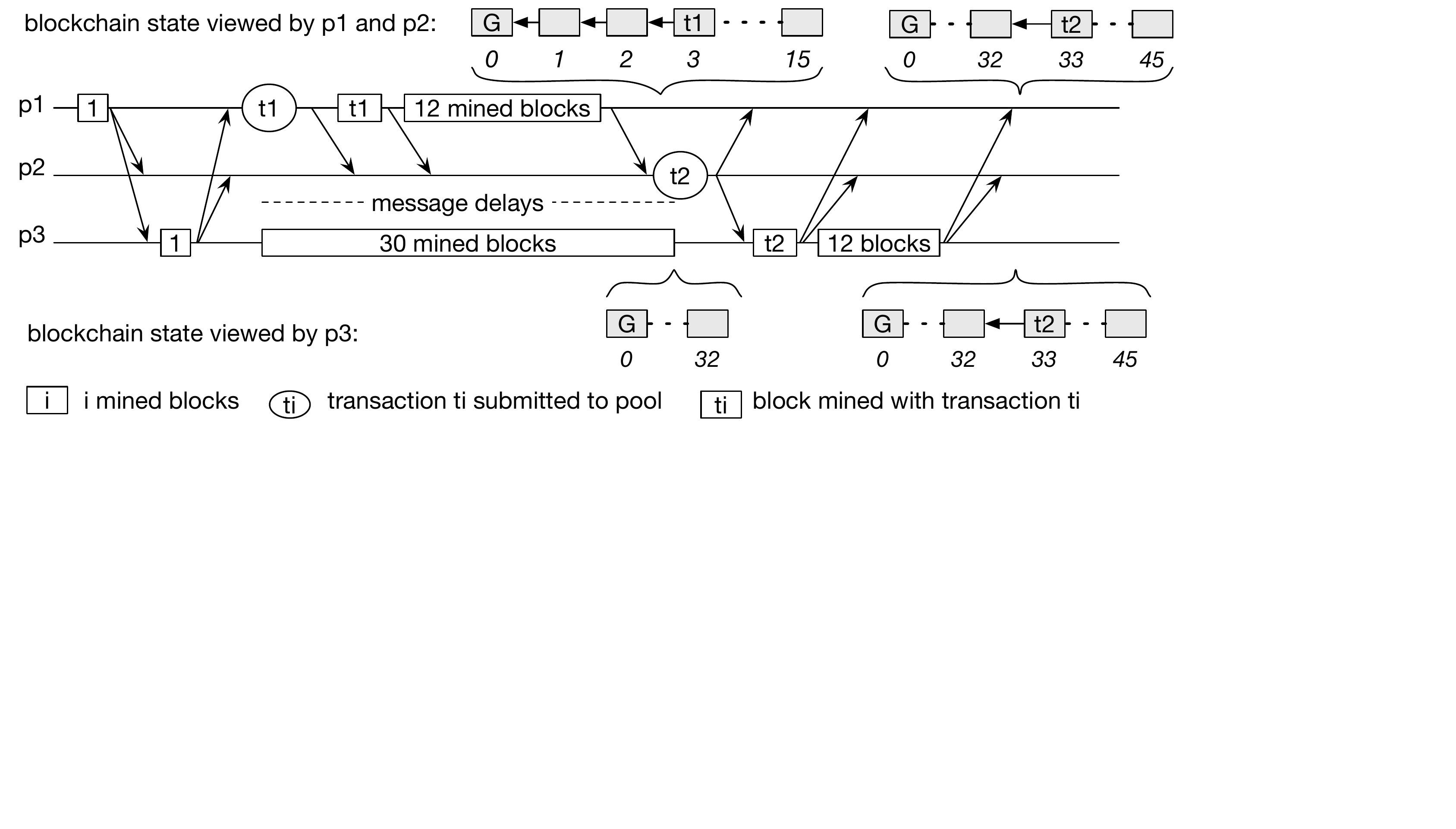}
\vspace{-3em}
\caption{Execution scenario leading to the Blockchain anomaly: $p_3$ mines a longer chain than $p_1$ without including $t_1$ and without disseminating new blocks until it forces a reorganisation that imposes $t_2$ to be committed while $t_1$ appears finally uncommitted\label{fig:execution}}
\end{figure*}

\subsection{Tracking Blockchain Anomalies}
Another dramatic aspect of the Blockchain anomaly is that it goes undetected.
%despite the transparency advantage of blockchains that capture events in an immutable storage that can be audited.
More specifically, the Blockchain anomaly relies on a wrongly committed state of the blockchain. Once the wrongly committed state gets uncommitted, there is no way to a posteriori observe this problematic state and to notice that a blockchain anomaly occurred.
Although it is possible to observe that a peer mined several blocks in a row, there is no way to track down the beneficiaries of the Blockchain anomaly.
%Not only are there no ways to observe the Blockchain anomaly, but there is no way to track down the beneficiaries of the Blockchain anomaly.
This dangerously incentivises participants of the private chain to leverage the Blockchain anomaly to attack the chain.

\section{Experimental Evaluation}\label{sec:eval}

In this section, we describe a distributed execution involving a private chain that 
results in the Blockchain anomaly.
%present our experiments resulting in the Blockchain anomaly.

\subsection{Experimental Setup}

% private chain in our local area network
We deployed a private blockchain system in our local area network using geth version 1.4.0, which is a Go implementation 
of the command line interface for running an Ethereum node.
%To this end, we used Ethereum that relies on the same promising proof-of-work concept as Bitcoin but offers an open source software with support for private chains.
%The experimentations ran on a closed local area network running a \textit{private chain} of Ethereum. 
We setup three machines connected through a 1\,Gbps network, two consisting of miners, $p_1$ and $p_3$, generating blocks and one consisting of a peer $p_2$ simply submitting transactions.
%\footnote{A peer is a client that is not mining but using the system for transactions} 
%which acted as a simple client that only sent transactions. 
Peers $p_1$ and $p_2$ consist of 2 machines with 4 $\times$ AMD Opteron 6378 16-core CPU running at 2.40\,GHz with 512\,GB DDR3 RAM, each.
Peer $p_3$ consists of a machine with 2 $\times$ 6-core Intel Xeon E5-260 running at 2.1\,GHz with 32\,GB DDR3 RAM.
%The experiment utilised three machines as follows: one miner and one peer having 4 $\times$ AMD Opteron 6378 16-core CPU @ 2.40GHz with 512GB DDR3 RAM each, and another miner having 2 $\times$ Intel Xeon E5-260, 2 $\times$ 6 cores  @ 2.1GHz,  with 32GB DDR3 RAM. 

%\paragraph{Connection delays}
%To illustrate the anomaly, 
We artificially created a network delay by transiently annihilating connection points between machines. 
Note that such artificial delays could be reproduced by simply unplugging an ethernet cable connecting a computer to the company network and does not require an employee to 
access physically a switch room. 

Also, we made sure $p_3$ would mine faster than $p_1$, by mining with the 24 hardware threads of $p_3$ and a single hardware thread of $p_1$. The same speed difference could be obtained between a loaded server and a  
server that does run any other service besides mining.
%Note that in a more realistic setting a machine mining in Ethereum with an AMD Radeon R9 280X would get a 25-fold speedup over a machine mining with an Intel Core i7.
Note that hardware characteristics may also help one machine mine faster than the rest of a private chain network. For example, 
a machine equipped with an AMD Radeon R9 290X would mine faster in Ethereum than a pool of 25 machines, each of them mining 
%a 25-fold speedup over a machine 
with an Intel Core i7.

% machines and network configuration
%All the machines were connected in the local network using a 1 Gbps network connection directly switched between the servers. 
%%The experiment was conducted in a closed local network, conducted over the span of a day.
%We setup two miners \emph{(\textbf{$p_1$} and \textbf{P3})} and one peer \emph{(\textbf{P2})} submitting transactions using geth version 1.4.0, which is a Go implementation 
%of the command line interface for running an Ethereum node.
%%The Ethereum source we used was \textbf{Geth}, a \emph{Go} implementation of the Ethereum system, version 1.4.0 which was highly suited to the experimental setup. 
%It enabled us to modify the genesis block and utilise for testing corner cases.
%Finally, we artificially created a network delay by transiently anihiliating connection points between machines.
%%The servers were running geth, which had two miners \emph{(\textbf{$p_1$} and \textbf{P3})} and one peer \emph{(\textbf{P2})} that was just submitting transactions. 
%The anomaly is detected when the client tries to send money depending on some existing actions. The client only sends money once the peer has a certain verified amount in his wallet. If and only if the peer has been shown that the money had transferred to his wallet and the transaction had been validated, the peer will then perform the second transaction. \\

\subsection{Distributed Execution}

The default case of the anomaly occurs with a conditional transaction. A peer in the system has a condition that it will only send money if %a \emph{certain amount of money is in its wallet}. 
some other peers transferred him some coins successfully.
As mentioned previously, for the transaction to be committed, there must be at least $k$ blocks mined 
%\vincent{Define minting as the successful process of producing coins/assets as opposed to mining as the process of trying to solve the crypto puzzle!} 
after the block containing the transaction. %\footnote{As explained in the Ethereum Yellow Paper by Gavin Wood}. ~\cite{Woo14}
In our experiment, the client only sends coins once the peer owns a verified amount of coins. % has a certain verified amount in his wallet. 
The peer performs a transaction $t_2$ only if it was shown by the system that the previous transaction $t_1$ had been committed and the money was successfully transferred to its wallet.
%If and only if the peer has been shown that the money had transferred to his wallet and the transaction had been validated, the peer will then perform the second transaction

%\begin{figure}[t]
%\centering
%\includegraphics[width=0.3\textwidth]{txswap_legend}
%\caption{Legend for Figure~\ref{fig:execution}\label{fig:legend}}
%\end{figure}

Figure~\ref{fig:execution} %and~\ref{fig:legend} 
%depicts the execution that leads to the anomaly where the block denoted `G' is the genesis block.
%To observe the anomaly, the execution in \emph{Figure 2} was followed. 
%The initial state of the system has all three participants ($p_1$, $p_2$, $p_3$) connected to each other via the network.
depicts the distributed execution leading to the Blockchain anomaly where $p_1$, $p_2$ and $p_3$ exchange information about the blockchain whose genesis block is denoted `G'.
\begin{enumerate}
\item Peer $p_1$ mines a first block after the genesis block and informs $p_2$ and $p_3$ to update their view of the blockchain state.
\item Peer $p_3$ mines a second block and informs $p_1$ and $p_2$ of this new block.
\item A network delay is introduced between peers $p_1$ and $p_2$ on the one hand, and peer $p_3$ on the other hand. 
\item Peer $p_1$ submits transaction $t_1$ and informs $p_2$ but fails to inform $p_3$ due to the network delay. In the meantime, peer $p_3$ starts mining a long series of 30 blocks.
\item Peer $p_1$ mines a block that includes transaction $t_1$ and mines 12 subsequent blocks; $p_1$ then informs  $p_2$ but not $p_3$ due to the network delay.
\item Peer $p_2$ receives the notification from $p_1$ that $t_1$ is committed because its block and $k$ subsequent blocks are mined; then $p_2$ decides to submit transaction $t_2$ that should only execute after $t_1$.
\item The network becomes responsive and $p_3$ who receives the information that $t_2$ is submitted, mined $t_2$ in a block along with 12 subsequent blocks.
\item Once peers $p_1$ and $p_2$ receive from $p_3$ the longest chain of $45$ blocks, they adopt this chain, discarding or postponing the blocks that were at indices $2$ to $15$, including the transaction $t_1$, of their chain.
\item All peers agree on the final chain of 45 blocks in which $t_2$ is committed and where $t_1$ is finally
not committed before $t_2$.
%
%\item $p_1$ mines a block and gain some coins. 
%\item A network delay is then introduced between $p_1$ and $p_2$ on one side and $p_3$ on the other side, as if $p_3$ was disconnected transiently from the rest of the system.
%\item Since $p_3$ is a miner, it continues to mine blocks on its own, assuming it is performing mining on the main chain, when in reality it is mining a forked chain. 
%\item $p_1$ then submits a transaction (\emph{$t_1$}) to an account on $p_2$, which then gets mined onto a block in the chain seen by $p_1$ and $p_2$. 
%\item $p_2$ then waits for the block to be decided
%%\footnote{In Ethereum, at least 6 blocks are mined above} 
%which validates that his wallet has the given amount.
%\item At this stage, $p_2$ sends a transaction (\emph{$t_2$}) to another account on $p_1$ but
%\item before the block has been mined, the connection between $p_3$ gets restored and there is now connection between all three peers. 
%\item This then begins the synchronisation between the blockchain and transaction pool. 
%\item Since $p_3$ had mined a larger chain, it gets chosen as the correct, main chain for all participants of the system. This point happens during the mining of blocks, in which the second transaction (\emph{$t_2$}) has been mined into a block and submitted to the chain, before the first transaction (\emph{$t_1$}), which was then mined into a block after $t_2$. 
\end{enumerate}

\begin{figure}[t]
\centering
\includegraphics[width=0.5\textwidth, clip=true, viewport=0 0 500 460]{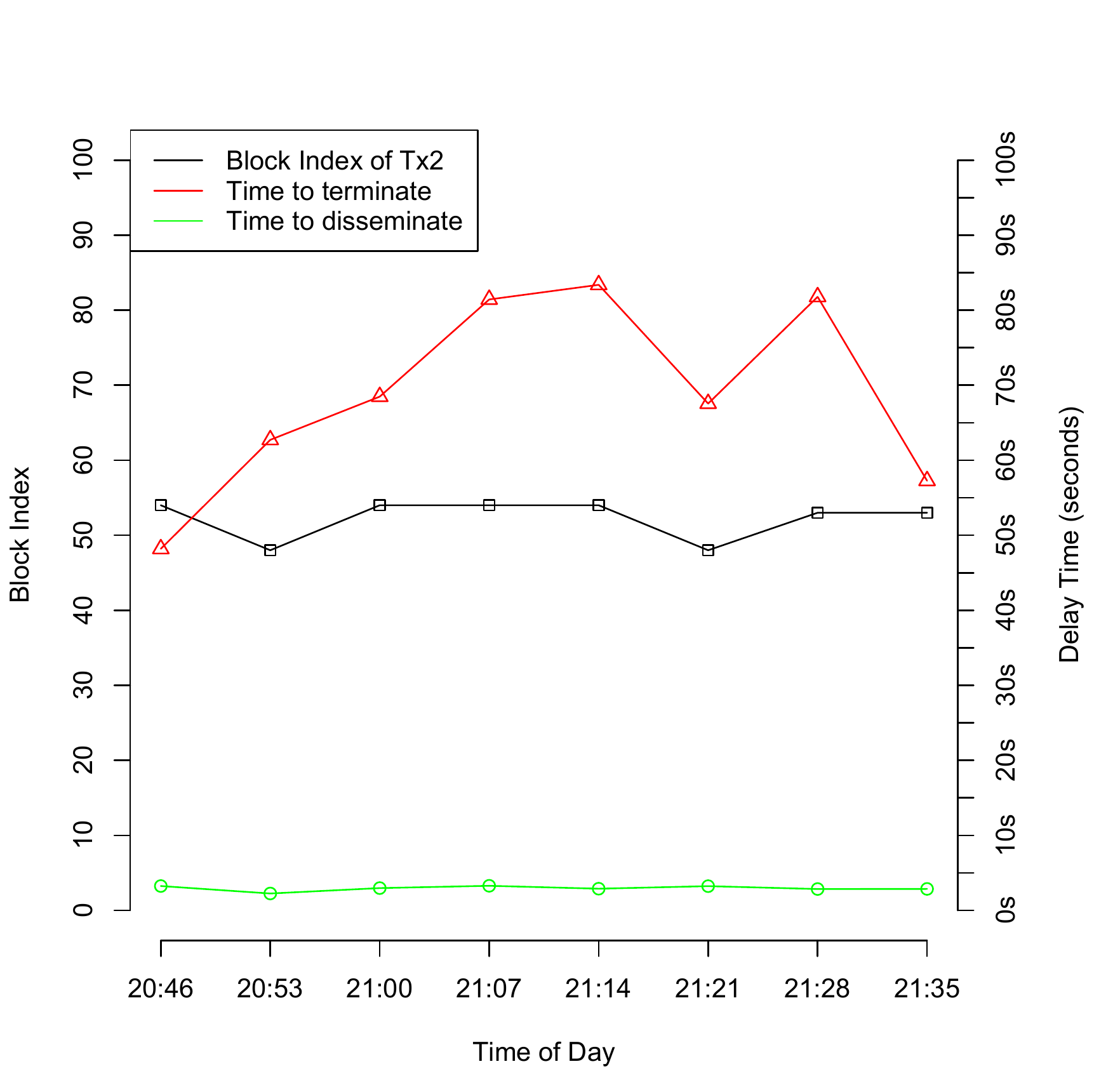}
\caption{Automated executions of the Blockchain anomalies over a period of $50\,min$, the execution is non-determinstic due to the randomness of the mining process and the network delay between peers\label{fig:eight_test}}
\end{figure}

% I can't remember what paper said the definitions of safety, termination etc. 
%This is a violation of the agreement property of consensus in Ethereum. 
This execution results in a violation of the conditional property of transaction $t_2$ stating that $t_2$ should only execute if $t_1$ executed first.
This violation occurred because transaction $t_1$ had been included in one chain, decided and agreed by two of the participants, it was then changed after the message of the third participant was finally delivered to the rest of the network.  
%\vincent{Can we say messages get finally delivered?}
%This has a crucial effect on the Ethereum system, as this presents the fact that termination was violated, and that may present potential problems. 

\subsection{Automating the Reproduction of the Anomaly}\label{sec:script}

To illustrate the anomaly, we wrote a script that automated the execution depicted in Figure~\ref{fig:execution}. 
Figure~\ref{fig:eight_test} represents the execution of a script that execute 8 iterations of the Blockchain anomaly over a period of 50 minutes. 
Again the goal is to wait until $t_1$ gets committed before issuing $t_2$ that ends up being committed while $t_1$ does not appear to be. 
Note that this is similar to Figure~\ref{fig:ba} except that $t_2$ is not necessarily included at the index $t_1$ occupied initially. In particular, the block in which $t_2$ gets included varies from one iteration to another due to the non-determinism of the execution as indicated by the curve with square points.
%
%The execution is non-deterministic as the index of the block in which $t_2$ gets committed varies from one iteration of the experiment to another. 
This non-determinism is explained by the randomness of the mining process and the latency of the network that also impacts the time it takes for the consensus to terminate (curve with triangle points) in each iteration of the experiment. Note that we use $k=11$ in this experiment, making sure that 12 blocks were successfully mined, as recommended since the release of Ethereum 1.3.5 Homestead, for the consensus to terminate.

As expected, in each of these eight cases we observed the Blockchain anomaly: even though $t_2$ was issued after $t_1$ was successfully observed as committed, if the messages get successfully delivered, then the reorganisation results in $t_2$ being committed while $t_1$ is not.  
Finally, we can observe that the time to disseminate 
a committed transaction to all the peers of the network is much shorter than the termination delay. 
This is due to the time needed to mine a block, which is significantly larger than the latency of our network.

\begin{figure}[t]
\centering
\includegraphics[width=0.5\textwidth, clip=true, viewport=0 0 500 460]{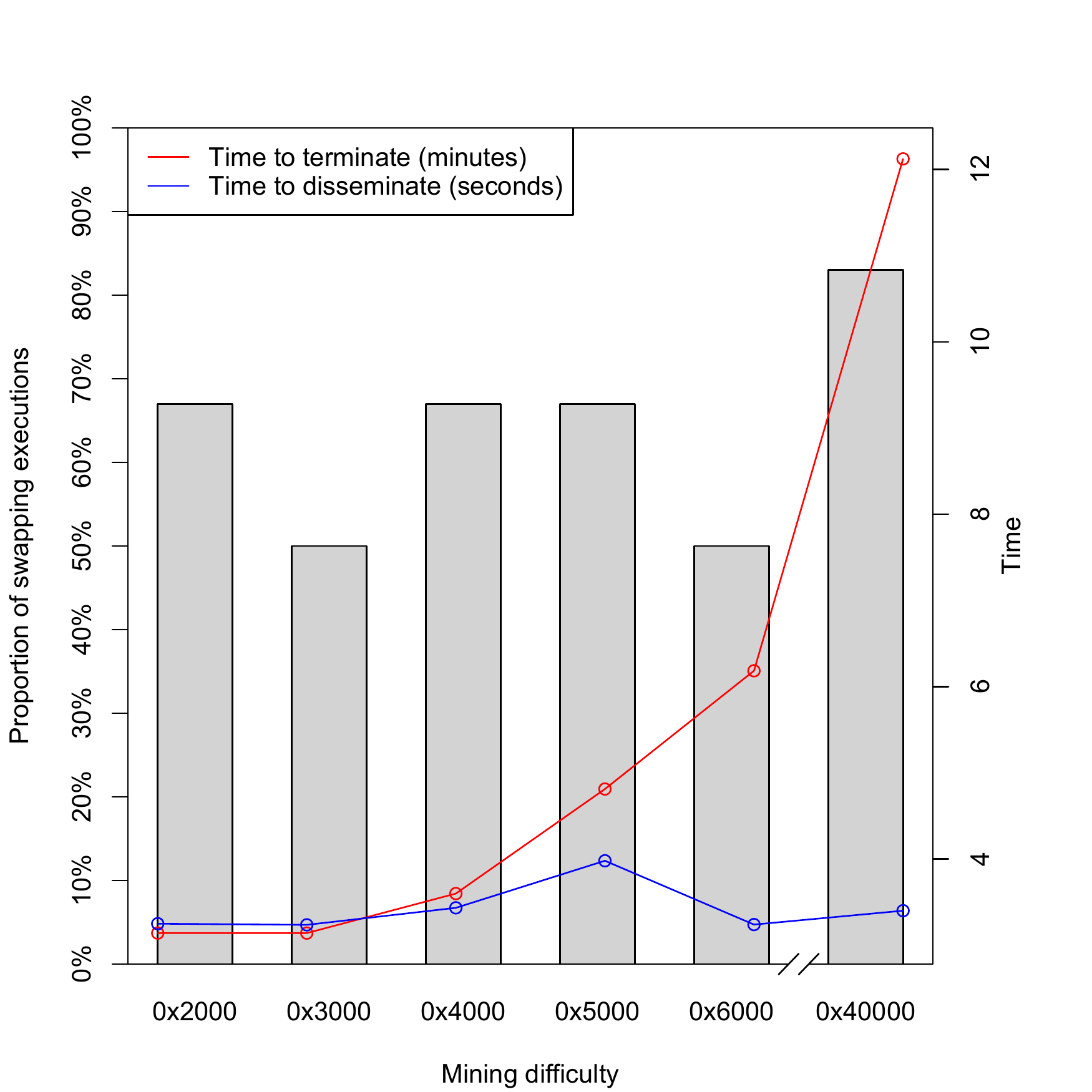}
\caption{The proportion of transaction swaps observed does not depend on the difficulty, as opposed to the consensus termination that increases with the difficulty\label{fig:k6}}
\end{figure}

\begin{figure*}
\begin{lstlisting}[language=JavaScript]
contract conditionalPayment {

	uint32 paid; // to keep track of the amount paid by Alice when deciding on Bob's transfer
	mapping (address => uint256) public balances; // map addresses to their respective balance
	address A = 0x57ec7927841e2d25aad5f335e3b701369b177392; // the address of Alice's account 
	address B = 0x5ae58375c89896b09045de349289af9034902905; // the address of Bob's account 

	modifier onlyFrom(address _address) {  // enables execution of functions depending on invoker
		if (msg.sender != _address) throw; 
		_ 
	}

	function sendTo(address B, uint32 _amount) onlyFrom(A) { // Alice sends money to Bob
		if (balances[A] >= _amount) {  // checking the sufficiency of funds available
			balances[A] -= _amount;
			balances[B] += _amount;
			paid = _amount;  // sorting the amount paid
		}
	}
	
	function sendIfReceived(address C, uint32 _amount) onlyFrom(B) { // Bob sends money to Carole
		if (paid > _amount) { // only if the previous payment was sufficient
			balances[B] -= _amount;
			balances[C] += _amount;
		} else {
			throw; // cancel contract execution
		}
	}
}
\end{lstlisting}
\caption{A smart contract written in the Solidity programming language to replace transactions prone to the blockchain anomaly: the \texttt{sendIfReceived} function checks that the transfer from A to B occurred before executing the transfer from B to C\label{fig:sc2}}
\end{figure*}

\subsection{Swap Frequency with Different Mining Difficulties}

%The anomaly was reproduced consistently several times with the experimental setup, using a basic genesis block and the default Ethereum difficulty. 
In the previous experiment, we used the default Ethereum difficulty (0x400) and automated the execution with a precise script. To better understand the cause of the anomaly we tried reproducing the anomaly by hand (without the script) with larger difficulties.

Figure~\ref{fig:k6} depicts the average number of blockchain anomalies leading to a swap (where both $t_2$ and $t_1$ are eventually committed in reverse order) occurring in our private chain for 6 different mining difficulties.
Each bar results from the average number of anomalies observed during 6 manual runs of the scenario depicted in Figure~\ref{fig:execution}. More precisely, the figure reports the \emph{swap} scenarios, where the first transaction $t_1$ gets successfully committed before $t_2$ gets issued, and eventually transactions $t_1$ and $t_2$ appear committed in reverse order.

%The first observation is that we could not reproduce the anomaly in all cases but observe it 6 times out of 10 executions.
%At first, we hypothesised that the difficulty was directly related to the occurrence of the Blockchain anomaly, that as the difficulty increased the swap would occur less due to the time taken to produce the hash and mine the block onto the chain. As we increased the difficulty, and emulated times beyond the normal chain\footnote{Chain time was 12.7 seconds per block} on Ethereum, we still found that the swap was occurring on average \emph{60\%} of the time.

We ran this particular experiment with $k=10$
%As we did our experiments before the release of Homestead, we used $k=10$ 
for the termination of consensus, meaning that $t_1$ was mined in block at index $i$ and it was committed once the chain depth reached $i+10$ blocks. (We presented the anomaly in the case where $k=11$ in Section~\ref{sec:script}.)
At first, we thought that the occurrence of this Blockchain anomaly was dependent on the difficulty of mining a block: the faster a block 
could be mined, the more likely the anomaly would occur.
%As shown in Figure~\ref{fig:k6}, the chain swaps have an unpredicted percentage of occurrence. This went against initial thoughts of the swap only occurring at low difficulties. 
To validate this, we varied the mining difficulties from 0x2000 to 0x40000 and measure the frequency of the Blockchain anomaly 
over 6 executions for each difficulty value. We observe that there was no significant correlation between the difficulty and the 
occurrence of the anomaly and that in average we could observe the swap 6 times out of 10.

We also measured the time it would take for consensus to terminate in these scenarios (upper curve) and observed, as expected, that 
the termination time was proportional to the difficulty. This is explained by the fact that the difficulty impacts the time needed to mine a 
block, which in turn, impacts the time it takes to mine $k+1$ blocks for termination. In addition we report the time it would take for a 
transaction in a mined block to be disseminated to all the peers of the network (bottom curve) and observed that it was not related to the 
difficulty.
Finally, we observed that having a network delay greater than the time to mine was foundational to the observation of the anomaly. 
%As the figure shows, the time to verify the block ascended with the difficulty as expected, as well as this, the time for transaction to be disseminated\footnote{All peers having the block in the chain from when it was initially mined} was averaging a constant value close to \emph{3.4}. 
%This shows that to observe this anomaly, a network delay greater than the time to validate 

%contract chainSwap {
%    function AtoB(address to_add) returns (bool sufficient){
%        while(true){
%            if(msg.sender.balance > 1000000000000000000){
%                to_add.send(1000000000000000000);
%                return true;
%            } else {
%                return false;
%            }
%        }
%    }
%}
%\begin{figure}
%\begin{lstlisting}[language=JavaScript]
%contract faultyContract {
%
%	uint32 paid;
%	/* This creates an array with all balances */
%	mapping (address => uint256) public balances;
%			
%	function sendTo(address B, uint32 _amount){
%		if (msg.sender != A) throw;
%		if (balances[A] >= _amount) {
%			// A sends money to B
%			balances[A] -= _amount;
%			balances[B] += _amount;
%			paid = _amount;
%		}
%	}
%	
%	function isSent() returns (bool result){
%		return paid != 0;
%	}
%	
%	function sendIfReceived(address C){
%		// B sends money to C
%		balances[B] -= _amount;
%		balances[C] += _amount;
%	}
%}
%\end{lstlisting}
%\caption{Example of a smart contract that will suffer from the Blockchain anomaly\label{fig:sc1}}
%\end{figure}

\begin{figure*}
\begin{lstlisting}[language=JavaScript]
contract problematicConditionalPayment {
	...
	function checkPayment(address B, uint32 _amount) onlyFrom(B) constant returns (bool result) { 
		if (paid > _amount) { // check that Alice paid
			return true;
		} else throw;
	}
	
	function sendIfReceived(address C, uint32 _amount) onlyFrom(B) { // Bob sends money to Carole
		balances[B] -= _amount;
		balances[C] += _amount;
	}
}
\end{lstlisting}
\caption{Executing the transfer to Carole in a separate function may suffer from the Blockchain anomaly\label{fig:sc3}}
\end{figure*}

\section{Smart Contracts}\label{sec:sc}

Smart contracts are a foundational aspect of the \emph{Ethereum} system, as they are distributed code execution based on conditional aspects. The contracts can be programmed to allow for certain conditions to be met in order for the code to be executed. 
%This was an essential part of our discovery, as smart contracts could perform the required actions to be able to solve the anomaly observations.
What we found was that the anomaly prevention depended entirely on the programming of the smart contract. This means that if a smart contract was coded so that it did not properly check the condition that the first transaction had occurred, it would execute as normal, acting like a normal transaction and suffering from the anomaly. 
%The contract allows for conditional statements to be checked, and for each time that the contract is executed, the peer burns an amount of gas\footnote{Ethereum currency in \emph{Ether}} to execute code. 

\subsection{On-Chain vs. Off-Chain Computation}
In Figure~\ref{fig:sc2}, we illustrate the writing of a smart contract in the Solidity programming language with which we could not observe the anomaly. The key point is that 
the \texttt{sendIfReceived} function groups two steps: the check that the amount has been paid at Line 22 and the payment that 
results from this successful check at Lines 23 and 24. Because these two steps are executed on-chain, we know that one has to 
be necessarily true for the second to occur. 

However, if the two steps were parts of two separate functions of the contract, one checking 
that the amount had been paid and another that would do the payment and be invoked upon the returned value of the former then the 
anomaly could arise. For example, consider Figure~\ref{fig:sc3} where one function, \texttt{checkPayment}, checks that the payment from Alice proceeded correctly (Lines 3--7) and the other function, \texttt{sendIfReceived}, is modified to execute the payment unconditionally (Lines 9--13). Even if Bob invokes \texttt{checkPayment} and observes that it returns successfully before invoking \texttt{sendIfReceived} the anomaly may arise. The reason is that the check is made off-chain and nothing guarantees that the payment from Alice was not reorganised while Bob was checking the result off-line.

To conclude, it looks like the former contract in Figure~\ref{fig:sc2} has higher chances of not suffering from the Blockchain anomaly than the smart contract of Figure~\ref{fig:sc3} as it executes the check and the conditional transfer on-chain, however, this does not guarantee that the smart contract of Figure~\ref{fig:sc2} is immune to the blockchain anomaly. Further investigation is needed to prove it formally. In addition and just like transactions, smart contracts must be included in a block that gets mined and appended to the blockchain. Its inclusion into the blockchain even with $k$ subsequent blocks may suffer from a reordering as well, and lead to other kind of anomalies.

\subsection{Multisignatures and the Case of Bitcoin}

Even without the Turing-complete scripting language, there may be ways in Bitcoin to bypass the Blockchain anomaly. 
The idea is to change a conditional transaction into a joint payment that includes both the conditional transaction and the action enabling its condition. The idea of including the transaction and the action is similar to the idea of grouping in the same contract function \texttt{SendIfReceived} of Figure~\ref{fig:sc2}, the check and the transfer that we described before.

The joint payment will represent the payment of Carole by both Alice and Bob. The payment will thus take two inputs, owned by different people, and give one output. Because the coins of these two inputs come from different addresses, the joint payment needs two different signatures. The joint payment can be achieved with a \texttt{multisig} transaction in Bitcoin so that the \texttt{multisig} transaction requires either two signatures from Alice, Bob and an arbitrator, Donald, in order to execute. If both Alice and Bob sign the transaction, then it executes and Carole gets paid. However, if Alice or Bob refuses to sign, then Donald can help resolving the transaction by signing. It is important to note that the semantic of the joint payment differs from the conditional transaction though: Bob cannot wait until he gets the money from Alice to choose what to do, whether to pay Carole.

\section{Discussion}\label{sec:disc}

An interesting aspect of Nakamoto's consensus is that if the system is large enough and the mining power is sufficiently scattered among enough mining pools, then the probability of having a mining pool mining faster than the other can be made arbitrarily small. For this reason, the Blockchain anomaly has a very low chance of occurring in realistic executions of a large-scale permissionless blockchain systems like Bitcoin or Bitcoin-NG~\cite{EGSvR16}. Recent work has shown, however, that incentives exist for miners to not disclose the block they successfully mine in order to waste the mining efforts of others, making it possible for them to mine a longer chain of blocks than others. This could potentially lead in turn to the Blockchain anomaly~\cite{EG14}. 

The 51-percent attack, where an attacker who controls more than half of the mining power 
of the public network can mine blocks faster than others, could lead to the Blockchain anomaly in a public blockchain system.
The attacker can issue a transaction to convert some bitcoins to withdraw some money. 
Once the transaction is mined into a block at 
index $i$, then the attacker can fork the blockchain from index $i-1$, hence excluding his transaction, with a new series of blocks that 
gets eventually longer than the main chain. As the longest branch gets adopted, the attacker's transaction does not appear in the chain so that, in the end, the attacker withdrew some money while keeping his coins.
With the same technique, one could easily override the block containing the transaction from Alice to Bob.
%
%For example, the attacker that owns 51\% of the mining power can decide to undo a particular transaction in a block of index $i$ within a blockchain of depth $d$ by appending to block $i-1$ a branch containing more than $d-i+1$ consecutive blocks.
%For example
The possibility of such an attack was raised in the context of the Bitcoin public chain as the mining power was noted as insufficiently scattered to avoid coalition~\cite{EG14}.
%, this discussion is out of the scope of this paper.

%\vincent{Gun Sirer et al. showed that Bitcoin was vulnerable to attackers detaining more than $\frac{1}{4}$ of the mining power of the system and that miners had some incentives in keeping their block for themselves once found rather than discolsing it in the hope that other miners would waste their mining power.}

%An interesting attack to Ethereum that the Blockchain anomaly facilitates is the long-range attack, where the attacker forks the chain from an earlier block, potentially down to the genesis block and mines contracts that are computationally hard to validate but easy to produce. For example, using 

One may think that the blockchain anomaly is specific to proof-of-work as there exist blockchain systems not based on proof-of-work that would not suffer from this issue because they 
trade availability for consistency (as discussed in Section~\ref{sec:rw}). This is the case of some proof-of-stake blockchain systems, like Tendermint, that guarantees agreement and validity of consensus deterministically.
The blockchain anomaly however applies even to blockchain systems based on proof-of-stake. For example, Casper is a proof-of-stake alternative to the {\sc Ghost} reorganistion protocol used in Ehtereum. 
%Actually, there 
%The blockchain anomaly is not specific to proof-of-work
It looks like proof-of-stake does not necessarily solve the problem, as even Casper favours availability over consistency.\footnote{\url{http://ethereum.stackexchange.com/questions/332/what-is-the-difference-between-casper-and-tendermint/536}.}

%contract chainSwap {
%    uint64 txPos;
%    function chainSwap(){
%        txPos = 0;
%    }
%    
%    function AtoB(address to_add, uint amt) returns (bool sufficient){
%        if(txPos == 0){
%            txPos = 1;
%            if(msg.sender.balance > amt){
%                to_add.send(amt);
%                return true;
%            }
%        }
%        return false;
%    }
%    
%    function BtoC(address to_add) returns (bool sufficient){
%        if(txPos == 1){
%            // specific case of B to C as shown in the execution
%            if(msg.sender.balance > 1000000000000000000){
%                to_add.send(1000000000000000000);
%                return true;
%            }
%        }
%        return false;
%    }
%}

Another problem raised by Gavin Wood, one of the founder of Ethereum, indicates that reorganisation can impact the initial order of transactions. This matters in an execution where two transactions aim at transfering \$100 from the same account whose initial balance is only \$100 because only the transaction that is committed first can be executed.\footnote{\url{https://blog.ethereum.org/2015/08/08/chain-reorganisation-depth-expectations/}.} The Blockchain anomaly is more general than this problem, in particular 
%the smart contract represented in Figure~\ref{fig:sc3} addresses this problem but not the blockchain anomaly. 
the Blockchain anomaly allows conflicting transactions to be successfully executed and committed in two 
different states of the blockchain.
Because the blockchain anomaly is more general, solving the blockchain anomaly would also solve this problem.

As it is known to be impossible to solve consensus in an asynchronous system in the presence of failures, researchers generally consider that a protocol ensures termination or agreement deterministically but not both.  In this paper, we considered that the blockchain consensus terminates deterministically based on the recommended 6 to 12 mined blocks of Bitcoin~\cite{Nak08} and Ethereum~\cite{Woo15} but sometimes failing at ensuring agreement.
%, because some decided transactions suffered from the Blockchain anomaly. 
Note that other formalisations also consider that termination of Nakamoto's consensus is deterministic and that only its safety property is probabilistic~\cite{GKL15}, just like we did.
One may argue however that termination is not guaranteed deterministically but rather probabilistically and that one can increase the probability of consensus agreement by simply delaying the termination; the characterisation of Nakamoto's consensus in Bitcoin-NG adopts this definition~\cite{EGSvR16}. In practice, however, blockchain applications assume consensus termination to provide a responsive service, 
as explained in Section~\ref{ssec:termination}.
For example, Vitalik Buterin, one of the founder of Ethereum, explained that waiting for 12 mined blocks is probably sufficient for the first block to be irreversible.\footnote{\url{http://ethereum.stackexchange.com/questions/183/how-should-i-handle-blockchain-forks-in-my-dapp/203\#203}.}
This can be true in large-scale permissionless system where the mining power is sufficiently scattered among mining pools, but as the Blockchain anomaly shows, it is easy to revert it in a private chain context.

%One can argue that proof-of-work blockchain systems that do not guarantee termination deterministically, as one can always consider that if agreement is not reached  

\section{Related Work}\label{sec:rw}

Proof-of-work has been previously compared to Byzantine fault tolerant protocols~\cite{CDE16,Vuc16}.
Some of this research~\cite{CDE16} focuses on comparing experimentally Bitcoin against PBFT~\cite{CL02}.
The Bitcoin blockchain and the PBFT consensus protocol were evaluated with nodes scattered at 
8 locations around the world. As one could expect given the difficulty of the crypto puzzle of Bitcoin, the experiments
showed that PBFT achieves a lower latency and a higher throughput than Bitcoin in serving transactions.
However, PBFT suffers from scalability limitations and the authors recommend using sharding to avoid having to scale 
to hundreds of nodes.

Another part of this research~\cite{Vuc16} discusses the probabilistic guarantees of proof-of-work systems and the deterministic 
guarantees of Byzantine fault tolerance. The proof-of-work consensus is compared to Byzantine agreement protocols along 
two axes, scalability and performance, where proof-of-work consensus protocols are considered as scalable but inefficient 
while Byzantine agreement protocols are considered as efficient but not scalable.
For example, Bitcoin scales beyond 1000 nodes while achieving  a performance lower than 100 transactions per second with a high latency, 
whereas standard Byzantine fault tolerant protocols achieve more than 10,000 transactions per second but scale only to tens of nodes. 

Some solutions immune to the Blockchain anomaly also exist.
PeerCensus~\cite{DSW16} was proposed as an algorithm with two components: one to execute a Byzantine agreement protocol on top of Bitcoin with a simple voting system and another 
to minimise the effect of Sybil attacks during these votes. The latter component makes it difficult for an attacker to create multiple identities so as to outnumber the 
votes with its own votes. Using this technique PeerCensus strengthens the guarantees of Bitcoin and resolves immediately the forks, hence avoiding the 
Blockchain anomaly.

Tendermint\footnote{\url{http://tendermint.com}.} is a blockchain system building upon proof-of-stake. It is known to favour consistency over availability, taking the opposite view of Casper, the proof-of-stake alternative to the 
{\sc Ghost} protocol. The Tendermint consensus protocol builds upon the Byzantine agreement protocol with authentication~\cite{DLS88} and requires strictly more than two third of correct 
processes to ensure agreement and validity deterministically and to guarantee termination when the network stabilises and messages between non-faulty nodes get delivered.

Although the Paxos anomaly was not considered a problem in the original design of Paxos~\cite{Lam98}, this scenario was informally stated as an anomaly during the design of the Zookeeper distributed coordination service~\cite{HKJ+10}, due to the engineers needing to implement conditional concurrent requests: Zookeeper organises nodes into a tree structure and it was desirable for the additions of a parent node and its child to be made concurrent.
The child addition depended naturally on the success of the parent addition.
Note that for other applications that do not need concurrent dependent requests Paxos is sufficient~\cite{GBFS15}.
The Paxos anomaly differs from the Blockchain anomaly because it can occur on two transactions issued by the same client. In Blockchain systems, a timestamp can be used to order the 
transactions issued by the same client. Another major difference between the Paxos and the Blockchain anomalies is that if consensus is reached, the index of the decision cannot change while the Blockchain anomaly precisely stems from 
the fact that the index of a decided transaction, or the order of its block in the chain, can change.

\section{Conclusion}\label{sec:conclusion}

This paper presents the Blockchain anomaly. Named after the Paxos anomaly, it prevents a user of 
mainstream blockchain systems from executing a conditional transaction, a transaction that should only 
execute in the current observable committed state or a later state of the system. 
Our experience of the use of an Ethereum private chain at NICTA/Data61 revealed the easiness of reproducing the anomaly by reordering transactions after they had been committed.
%\vincent{Benchmarking. could we make the scripts available online? does it make sense?}
A possible way to avoid the anomaly could be to write smart contracts rather than transactions, yet it adds to the level of complexity.

Our conclusion is that blockchain systems are difficult to use properly. This observation should discourage users from using blockchain systems unless they fully understand the underlying design principles and the guarantees they offer.
If we combine the facts that blockchain applications 
require consensus to terminate fast 
while the underlying blockchain protocols guarantee agreement probabilistically then we can obtain
dramatic results when applied to private chains.

Besides the prominent blockchain systems we have discussed, namely Bitcoin and Ethereum, there exist many alternatives. Exploring the alternatives that exclusively offer deterministic guarantees for private chain are part of  future work. 

\subsection*{Acknowledgements}
We wish to thank our colleagues Alexander Ponomarev and Mark Staples for their feedback on an earlier version of this paper 
%discussions related to the behavior of Ethereum 
and for pointing out Gavin Wood's blog post on the insufficient fund problems of Ethereum.
NICTA is funded by the Australian Government through the Department of Communications and the Australian Research Council through the ICT Centre of Excellence Program.

\bibliographystyle{abbrv}
\bibliography{bib}

%\end{multicols}

\end{document}